\documentclass[pra,reprint,amsmath,amssymb,floatfix,citeautoscript,superscriptaddress]{revtex4-1}
\usepackage{cancel}
\usepackage{amsmath}
\usepackage{physics}
\usepackage{graphicx}
\usepackage{amssymb}
\usepackage{amsthm}
\usepackage{bm}
\usepackage{dcolumn}% Align table columns on decimal point
\usepackage{braket}% Enables braket notation
\usepackage{longtable}

\usepackage{ragged2e}
\usepackage{txfonts}

\usepackage{color}
\usepackage[usenames,dvipsnames]{xcolor}
\definecolor{myblue}{rgb}{0,0,1}
\usepackage[breaklinks=true,colorlinks=true,linkcolor=myblue,urlcolor=myblue,citecolor=myblue]{hyperref}

\newcommand{\vq}{{\bm{q}}}
\newcommand{\vk}{{\bm{k}}}

\newcommand{\eps}{{\varepsilon}}

\newcommand{\kFup}{{k_\mathrm{F\uparrow}}}

\begin{document}

\title{The normal state of attractive Fermi gases from coupled-cluster theory}

\author{James M. Callahan}
\affiliation{Department of Chemistry, Columbia University, New York, New York 10027 USA}
\author{John Sous}
\altaffiliation{Present address: Department of Physics, Stanford University, Stanford, CA 93405, USA}
\affiliation{Department of Physics, Columbia University, New York, New York 10027 USA}
\author{Timothy C. Berkelbach}
\email{tim.berkelbach@gmail.com}
\affiliation{Department of Chemistry, Columbia University, New York, New York 10027 USA}
\affiliation{Center for Computational Quantum Physics, Flatiron Institute, New York, New York 10010 USA}

\begin{abstract}
We introduce coupled-cluster (CC) theory for the numerical study of the normal
state of two-component, dilute Fermi gases with attractive, short-range
interactions at zero temperature. We focus on CC theory with double excitations
(CCD) and discuss its close relationship with---and improvement upon---the
$t$-matrix approximation, i.e., the resummation of ladder diagrams via a
random-phase approximation.  We further discuss its relationship with Chevy's
variational wavefunction ansatz for the Fermi polaron and argue that CCD is its
natural extension to nonzero minority species concentrations. Studying normal state
energetics for a range of interaction strengths below and above unitarity, we
find that CCD yields good agreement with fixed-node diffusion Monte Carlo. 
We find that CCD does not converge for small polarizations and large interaction
strengths, which we speculatively attribute to the nascent instability to a superfluid state. 
\end{abstract}

\maketitle

\section{Introduction}

Two-component Fermi gases with tunable, attractive interactions exhibit a rich
phase diagram~\cite{ Giorgini2008,Randeria2014,Massignan2014,Zwierlein2014}.
Experimental realizations via ultracold atoms have enabled precision studies of
their quantum many-body physics, including fermionic superfluidity~\cite{
Kinast2004,Kinast2005,Zwierlein2005,Zwierlein2006,Shin2006,Partridge2006,Shin2008,Horikoshi2010,Riedl2011,Ku2012,Levin2012,Fenech2016,Mukherjee2017,Carcy2019,Sobirey2021}
and the smooth crossover from the Bardeen-Cooper-Schrieffer (BCS) limit at weak
attraction to the Bose-Einstein condensate (BEC) limit at strong
attraction~\cite{Greiner2003,Zwierlein2003,Partridge2005,Bartenstein2004,Regal2004,Zwierlein2004,Bourdel2004,Zwierlein2006a,Bloch2008,Stewart2008,Schirotzek2009,Navon2010,Feld2011,Bloch2012,Sommer2012,Koschorreck2012,Strinati2018,Ness2020}.
An ongoing investigation concerns the fate of the paired superfluid in the
presence of a spin polarization, with implications for superconductivity in
solids.  Specifically, at large polarizations, the superfluid is expected to
exhibit an instability to a partially polarized or fully polarized normal
state~\cite{
Bulgac2007,Schirotzek2008,Pilati2008,Punk2009,Bertaina2009,Chevy2010,Radzihovsky2010,Gezerlis2012,Ries2015,Mueller2017,Caldas2020,Parish2021}.

Despite the dilute nature of these systems, their precise phase boundaries and
related properties can only be determined by accurate quantum many-body
calculations.  Near unitarity, the strength of the interaction precludes simple
perturbative treatments~\cite{Ho2004,Nascimbene2010}.  This motivates the search
for affordable but accurate nonperturbative techniques, the most successful of
which have been quantum Monte Carlo methods, including diffusion Monte Carlo
(DMC)~\cite{
Lobo2006,Pilati2008,Bertaina2011,Astrakharchik2014,Galea2016,Galea2017,Schonenberg2017},
auxiliary field quantum Monte Carlo (AFQMC)~\cite{
Carlson2011,Shi2015,Bour2015,Vitali2017,Jensen2020,Jensen2020a}, diagrammatic
Monte Carlo~\cite{
Prokofev2008a,Burovski2008,Vlietinck2013,Kroiss2014,Kroiss2015,Goulko2016,VanHoucke2019,VanHoucke2020},
and others~\cite{Chang2004,Kolodrubetz2012,Anderson2015}.  Each of these methods
has its own limitations due to the fermion sign problem, e.g., necessitating the
fixed-node approximation in DMC or the phaseless approximation in AFQMC for
nonzero polarizations.  Moreover, calculating dynamical response functions via
QMC is an open challenge.

Here, we introduce coupled-cluster (CC) theory~\cite{Bishop1991}, which has been
successfully used in nuclear physics~\cite{Coester1960,Hagen2014} and quantum
chemistry~\cite{Crawford2000,Bartlett2007,Shavitt2009}, as a promising numerical
technique for the simulation of Fermi gases and related systems of cold atoms.
CC techniques are nonperturbative, systematically improvable, non-stochastic,
and sign-problem free.  Their computational cost scales polynomially with the
number of particles and orbitals.  In this work, we focus on CC theory with
double excitations as applied to the normal state at zero temperature.  This
version of CC theory has a number of important physical properties that support
its application to dilute Fermi gases: it is exact for interacting two-particle
problems, it is exact for noninteracting ensembles of interacting two-particle
problems (so-called size-consistency)~\cite{ Zhang2019}, and it fully includes
all ladder diagrams, which are known to dominate the physics of dilute systems
with strong, short-range interactions~\cite{
Bishop1973,Fetter2003,Shepherd2014}.

The layout of this article is as follows.  In Sec.~\ref{sec:theory}, we
introduce the theory underlying our study, including the Hamiltonian,
variational wavefunctions, CC theory, and the random-phase approximation.  In
Sec.~\ref{sec:results}, we present our results for the energies, Tan's contact,
and phase boundaries, as functions of polarization and interaction strength.
Finally, in Sec.~\ref{sec:conc}, we conclude by summarizing our work and
identifying many avenues for future studies.

\section{Theory}
\label{sec:theory}

In the low-density limit where the average interparticle separation is much
larger than the interaction range, the low-energy properties of Fermi gases are
determined by $s$-wave scattering and the interaction is completely specified by
the scattering length $a$.  We perform our study at zero temperature in three
dimensions with periodic boundary conditions in a cubic box of volume $V$
according to the Hamiltonian~\cite{Chevy2006}
\begin{equation}
\label{eq:ham}
\hat{H} = \sum_{\vk,\sigma} \eps_\vk \hat{a}^\dagger_{\vk,\sigma} \hat{a}_{\vk,\sigma}
    + \frac{g}{V} \sum_{\vk,\vk',\vq}
        \hat{a}_{\vk,\uparrow}^\dagger \hat{a}_{\vk',\downarrow}^\dagger 
        \hat{a}_{\vk'+\vq,\downarrow}^{} \hat{a}_{\vk-\vq,\uparrow}^{}
\end{equation}
where $\eps_\vk = k^2/(2m)$, $\hat{a}_{\vk,\sigma}^{}$
($\hat{a}_{\vk,\sigma}^\dagger$) annihilates (creates) a fermion with spin
$\sigma$ and wavevector $\vk$, and we restrict our study to the case of equal
masses $m$ (and set $\hbar=1$).  We discretize space into unit cells of volume
$b^3$ and sample the Brillouin zone using a uniform mesh of $N_k$ $k$-points,
where $N_k = V/b^3$.  The two-body interaction strength $g<0$ is chosen to have
the scattering length $a$ on the lattice,
\begin{equation}
g^{-1} = \frac{m}{4\pi a} - \int_\mathrm{BZ} \frac{d^3k}{(2\pi)^{3}}\frac{1}{2\varepsilon_\vk}
    = \frac{m}{4\pi a} - \frac{m\mathcal{K}}{4\pi b},
\end{equation}
where $\mathcal{K} = 2.442\ 749$ for the quadratic dispersion used here~\cite{
Mora2003,Pricoupenko2007,Ketterle2008REAL,Werner2012}.  Continuum results are
obtained in the limit $N_k \rightarrow \infty$ with the particle number fixed.

We consider a partially polarized Fermi gas containing $N_\uparrow$ majority
particles and $N_\downarrow$ minority particles in the box of volume $V$.  In
the above limits, the properties of the Fermi gas are universal and defined only
by the dimensionless interaction strength $1/\kFup a$~\cite{
Ho2004,Randeria2014}, where $\kFup$ is the Fermi wavevector of the majority
particles.  On the basis of the behavior at low minority spin concentrations
$x=N_\downarrow/N_\uparrow$, the ground-state energy of the normal state has
been parameterized by the form~\cite{ Lobo2006,Pilati2008,Mora2010}
\begin{equation}
E = \frac{3}{5} N_\uparrow \varepsilon_{\mathrm{F\uparrow}} \left(1-Ax+\frac{m}{m^\ast}x^{5/3}+F x^2\right),
\label{eq:eos}
\end{equation}
where $\varepsilon_{\mathrm{F\uparrow}}$ is the Fermi energy of the majority
particles and the parameters $A$, $m^\ast$, and $F$ depend on the interaction
strength $1/\kFup a$.  Specifically, $A$ and $m^\ast$ are the polaron binding
energy and effective mass, respectively, while $F$ quantifies the interactions
between polaron quasiparticles.

Using a single determinant corresponding to a filled Fermi sea $|0\rangle$
yields the energy
\begin{equation}
E_0 = \langle 0|\hat{H}|0\rangle 
    = \frac{3}{5} N_\uparrow \varepsilon_{\mathrm{F\uparrow}} \left(1 + x^{5/3}\right) 
        + \frac{g}{V} N_\uparrow N_\downarrow,
\end{equation}
which is the sum of the non-interacting energy and the mean-field interaction
energy; the latter vanishes in the continuum limit $N_k\rightarrow \infty$.
Thus the non-interacting and mean-field theories predict $A=F=0$ and $m^\ast=m$.

Interaction effects can be accounted for by using a linear combination of
determinants that are defined with respect to the filled Fermi sea.  Such
variational wavefunctions have been extensively used in the study of Fermi
gases~\cite{ Mathy2011,Trefzger2012,Parish2013,Cui2020}.  Here we will consider
the method of configuration interaction with double excitations (CID), with
ground-state wavefunction $|\Psi_0\rangle = (c_0+\hat{C}_2)|0\rangle$, where
\begin{subequations}
\begin{align}
\hat{C}_2 &= \sum_{\vk_\uparrow,\vk_\downarrow\vq}^\prime c_{\vk_\uparrow\vk_\downarrow\vq} 
    \hat{A}^\dagger_{\vk_\uparrow\vk_\downarrow\vq}, \\
\hat{A}^\dagger_{\vk_\uparrow\vk_\downarrow\vq} &=
    \hat{a}^\dagger_{\vk_\uparrow+\vq,\uparrow} \hat{a}^\dagger_{\vk_\downarrow-\vq,\downarrow} 
    \hat{a}_{\vk_\downarrow,\downarrow}^{} \hat{a}_{\vk_\uparrow,\uparrow}^{},
\end{align}
\end{subequations}
and the primed summation requires that $k_\sigma<k_{\mathrm{F}\sigma}$,
$|\vk_\uparrow+\vq|>\kFup$, and $|\vk_\downarrow-\vq|>k_{\mathrm{F}\downarrow}$.
The $\hat{C}_2$ operator thus creates all possible momentum-conserving double
excitations (two particles and two holes), with one excitation for each spin
type.  Although CID (and CCD as we discuss below) can also include same-spin
double excitations, our numerical testing showed no significant difference and
so we exclude them in all results that follow.  For a normalized wavefunction,
the variational CID energy is
\begin{equation}
\label{eq:ci_energy}
E = \langle \Psi_0|\hat{H}|\Psi_0\rangle
    = E_0 + \frac{g}{V}\sum_{\vk_\uparrow\vk_\downarrow\vq}^{\prime} c_{\vk_\uparrow\vk_\downarrow\vq}.
\end{equation}

\begin{figure*}[t!]
  \centering
    \includegraphics[scale=0.9]{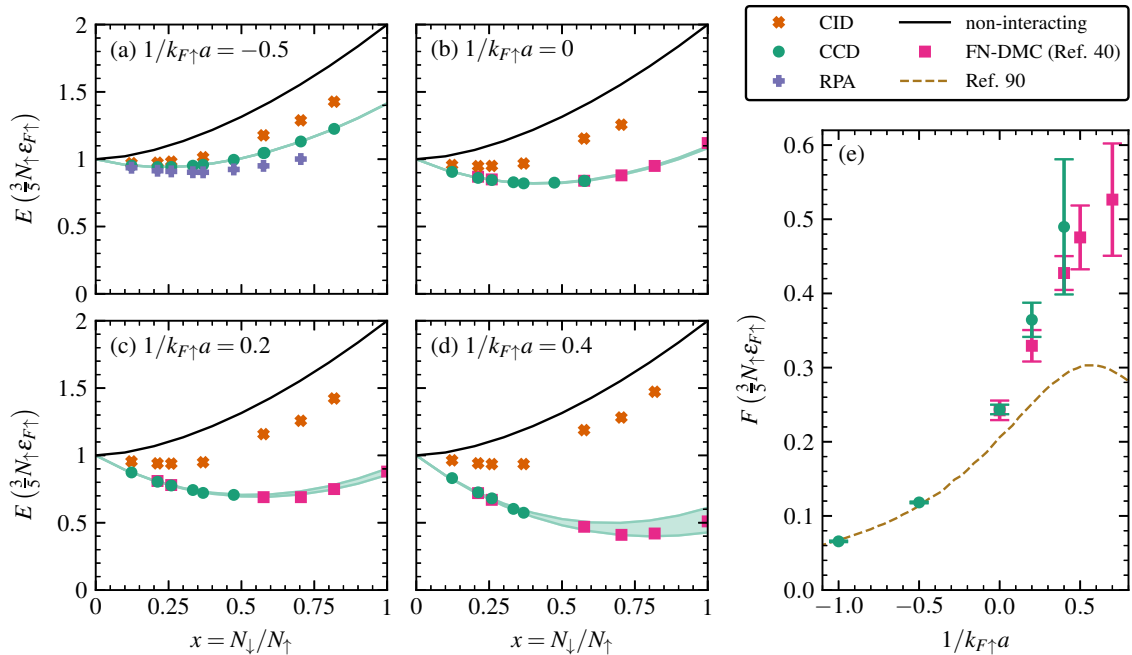}
  \caption{
  Energetics of the normal state of attractive Fermi gases over a range of spin polarizations and interaction strengths.
  (a)-(d) Total energy (scaled by the non-interacting energy of the spin-up particles)
  as a function of the concentration of minority spin-down particles
  at respective interaction strengths of $1/\kFup a = -0.5, 0, 0.2, 0.4$.
  Our continuum predictions from CID (crosses), CCD (circles), and ppRPA (plusses)
  are compared with FN-DMC results (squares) from Ref.~\onlinecite{Pilati2008}.
  As a guide, solid black lines give the total energy in the absence of two-body interactions.
  The green shaded region indicates the range of one-parameter fits for the interaction parameter $F$
  using Eq.~\ref{eq:eos}, as detailed in the text.
  (e) Comparison between $F$ calculated for CCD and FN-DMC results,
  compared to the analytic result of Ref.~\onlinecite{Mora2010}.
  Error bars indicate two standard deviations of the $F$ parameter fit.
}
  \label{fig:energies}
\end{figure*}

In the limit of a single minority atom interacting with a sea of majority atoms
(i.e., the Fermi polaron problem), the CID wavefunction above is the same as
Chevy's ansatz~\cite{Chevy2006,Combescot2007}, which provides remarkably
accurate energetics even in the limit of strong interactions.  For example, at
unitarity, Chevy's ansatz gives a polaron energy of
$E_\mathrm{p} \approx -0.6066 \varepsilon_{\mathrm{F}\uparrow}$,
which is extremely close to the diagrammatic Monte Carlo energy of 
$E_\mathrm{p} \approx -0.6157 \varepsilon_{\mathrm{F}\uparrow}$~\cite{Vlietinck2013,VanHoucke2020}.
Configuration interaction calculations are systematically improvable by
considering additional particle-hole excitations, as was done in
Ref.~\onlinecite{ Combescot2008}.  In quantum chemistry, such calculations would
be described as configuration interaction with double, triple, quadruple, etc.
excitations (CIDT, CIDTQ, etc.).

In principle, such variational wavefunctions can be straightforwardly applied to
partially polarized Fermi gases with nonzero minority spin concentrations $x$,
as studied here.  However, as is well-known in the quantum chemistry community,
such variational wavefunctions lack the important property of size-extensivity,
$E(N) \propto N$, or the closely related property of size-consistency, which can
be traced to the inclusion of unlinked diagrams in a diagrammatic expansion of
their total energies~\cite{ Bartlett1977,Bartlett1978,Bartlett2007,Shavitt2009}.
Therefore, energies obtained from truncated CI expansions are expected to
deteriorate for systems with increasing numbers of particles.  Unfortunately,
this behavior makes it inappropriate to extend Chevy's simple but successful
wavefunction ansatz to partially polarized Fermi gases, which we demonstrate
numerically in this work (\textit{vide infra}).

One of the most successful size-extensive theories of quantum many-body systems
is coupled-cluster theory, which can be truncated after any number of
particle-hole excitations, analogous to the variational CI wavefunctions
discussed above.  For example, the (right-hand) wavefunction used in
coupled-cluster theory with double excitations (CCD) 
is $|\Psi_0\rangle = e^{\hat{T}_2} |0\rangle$, where
\begin{equation}
\hat{T}_2 = \sum_{\vk_\uparrow,\vk_\downarrow\vq}^\prime t_{\vk_\uparrow\vk_\downarrow\vq} 
    \hat{A}^\dagger_{\vk_\uparrow\vk_\downarrow\vq}
\end{equation}
The amplitudes $t_{\vk_\uparrow\vk_\downarrow\vq}$ are determined to satisfy the nonlinear CCD equations
\begin{equation}
\label{eq:t2eqns}
0 = \langle 0| \hat{A}_{\vk_\uparrow\vk_\downarrow\vq} e^{-\hat{T}_2} \hat{H} e^{\hat{T}_2} |0\rangle
\end{equation}
from which the energy is evaluated as
\begin{equation}
\label{eq:cc_energy}
E = \langle 0|e^{-\hat{T}_2}\hat{H}e^{\hat{T}_2}|0\rangle 
    = E_0 + \frac{g}{V}\sum_{\vk_\uparrow\vk_\downarrow\vq}^{\prime} t_{\vk_\uparrow\vk_\downarrow\vq}.
\end{equation}
For the uniform systems studied here, the memory cost of 
CCD is $O(N_\uparrow N_\downarrow N_k)$ and the computational cost of iteratively
solving Eqs.~(\ref{eq:t2eqns}) is $O(N_\uparrow N_\downarrow N_k^2)$.

The CCD energy~(\ref{eq:cc_energy}) is nonvariational because $\hat{T}_2^\dagger \neq -\hat{T}_2$.
Rather, CCD is an infinite order perturbation theory.
To lowest order in $g$, the $t$-amplitudes are given by
\begin{equation}
t_{\vk_\uparrow\vk_\downarrow\vq} = \frac{g}{V} 
    \left(\varepsilon_{\vk_\uparrow+\vq}+\varepsilon_{\vk_\downarrow-\vq}
    -\varepsilon_{\vk_\downarrow}-\varepsilon_{\vk_\uparrow}\right)^{-1},
\end{equation}
which, together with Eq.~(\ref{eq:cc_energy}), just gives the second-order perturbation theory
energy. This perturbative energy expression was used to provide a picture of interacting Fermi polarons
in Ref.~\onlinecite{Mora2010}. 

The full CCD energy contains all ring diagrams, ladder diagrams, and mixtures of
rings and ladders---all with their exchange counterparts, resulting in a
properly fermionic theory~\cite{Shepherd2014}.  Ladder diagrams are known to be
important at low densities and such a restricted theory is equivalent to a
non-self-consistent $t$-matrix approximation, also known as the
particle-particle random-phase approximation (ppRPA)~\cite{
Bishop1973,Fetter2003,Scuseria2013,Peng2013,Berkelbach2018}.
This ppRPA energy can be obtained as an approximation to CCD, by only including
a subset of the terms in Eqs.~(\ref{eq:t2eqns}).  Such $t$-matrix or ppRPA
approaches have been extensively used in the study of dilute Fermi gases~\cite{
Jin2010,Urban2014,Hu2018,Tajima2018,Mulkerin2019,Pini2019,Durel2020,Hu2022},
but their quantitative accuracy is largely confined to the regime of very weak
attractions.  In the next section, we show that the richer diagrammatic content
of CCD leads to results with significantly greater quantitative accuracy.

\section{Results}
\label{sec:results}

\subsection{Ground-State Energy}
We have applied CCD, CID, and ppRPA to the Hamiltonian~(\ref{eq:ham}) for
several values of $x=N_\downarrow/N_\uparrow$. To avoid breaking spatial
symmetries in the Fermi sea determinant, we limit our calculations to closed
shell configurations with $N_\sigma = 1,7,19,27,33$.  To obtain continuum
predictions, we performed calculations using two different mesh sizes $N_k$
and extrapolated the energies to $N_k\rightarrow\infty$, assuming $N_k^{-1/3}$
convergence~\cite{Burovski2006,Werner2012}.  For CID, we used $N_k=8^3, 10^3$;
while for CCD and ppRPA, we used $N_k=10^3, 12^3$.

The Fermi polaron problem is defined by the choice $N_\downarrow = 1$ and
$N_\uparrow \rightarrow \infty$, i.e., $x\rightarrow 0$.  For this problem, with
all values of $N_\uparrow$ and $N_k$, we find that CCD and CID give numerically
identical results, which are therefore identical to those of Chevy's
ansatz~\cite{Chevy2006,Combescot2007} in the appropriate limits.  Despite their
different physical content, CID and CCD are equivalent for the polaron problem:
neither include triple excitations involving one spin-down and two spin-up
particles, which are responsible for the majority of the energy difference
between Chevy's ansatz and diagrammatic Monte Carlo
results~\cite{Combescot2008}.  At polarizations away from the polaron limit, the
equivalence between CID and CCD no longer holds, which will be the main focus of
our work.

In Figs.~\ref{fig:energies}(a)-(d), we show the total energies as a function of
$x$ for four values of the interaction strength, $1/\kFup a = -0.5, 0, 0.2, 0.4$, 
where $1/\kFup a=0$ corresponds to unitarity.  For the latter three values
of the interaction strength, we also compared to fixed-node DMC from
Ref.~\onlinecite{Pilati2008}.  Although the CCD procedure does not always
converge at large $g$ or $x$ (see below), our CCD results agree very well with
DMC results.  In contrast, the CID results do not, and overall they show a
significantly smaller correlation energy, consistent with the method's lack of
size-extensivity.

We fit our CCD results to the Landau-Pomeranchuk equation of
state~(\ref{eq:eos}).  Because $A$ and $m^\ast$ can be extracted from the
polaron problem, for which CCD gives results identical to Chevy's ansatz, we use
the latter's parameters~\cite{Combescot2007,Vlietinck2013} and have only the
interaction parameter $F$ as a free parameter to be fit.  In
Fig.~\ref{fig:energies}(e), we show the extracted value of $F$ over the full
range of interaction strengths we studied.  Since the overall fit is relatively
insensitive to $F$, especially for lower values of $x$ where CCD can converge
more easily, we include error bars indicating two standard deviations from the
fit value.  The resulting range of fitted energy curves are plotted in
Figs.~\ref{fig:energies}(a)-(d) as shaded regions along with the raw data.  We
have also carried out an identical fitting procedure to the FN-DMC results from
Ref.~\onlinecite{Pilati2008} with $x<1$ using the same values of $A$ and
$m^\ast$ for consistency.  Within error bars, we find very good agreement
between CCD and FN-DMC, indicating that both provide an accurate description of
quasiparticle interactions.  Finally, in Fig.~\ref{fig:energies}(e), we include
results from Ref.~\onlinecite{Mora2010}, which argued that the interaction
parameter $F$ can be obtained from polaron properties via 
$F = (5/9)(d\mu/d\varepsilon_\mathrm{F\uparrow})^2$, where $\mu$ is the chemical
potential of a single polaron.  The agreement is good, especially in the weakly
interacting BCS limit $1/\kFup\rightarrow-\infty$ that is accurately described
by lowest-order perturbation theory.

\begin{figure}[t!]
  \centering
    \includegraphics[scale=0.9]{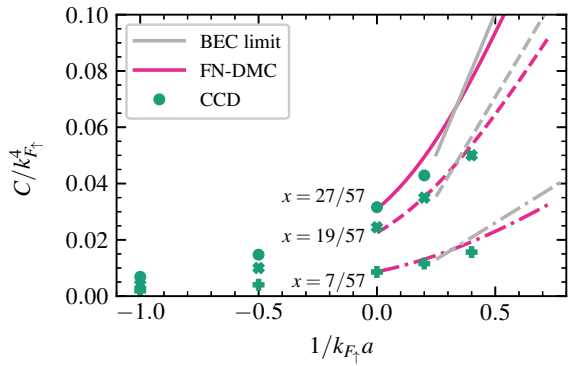}
  \caption{
  Universal contact as a function of the interaction parameter $1/\kFup a$ at
$x=27/57, 19/57, 7/57$ (top-to-bottom).  Grey (solid, dashed, and dashed-dotted,
top-to-bottom) lines on the right give, to leading order in $\kFup a$, the BEC
limit (at $a \rightarrow 0^{+}$).  Pink (solid, dashed, and dashed-dotted,
top-to-bottom) lines come from the analytic, parametrized formulas of
Ref.~\onlinecite{Bertaina2009} that were constructed to match FN-DMC results.
Green circles, crosses, and plusses (top-to-bottom) are calculated from our CCD
data as described in the main text.
}
  \label{fig:tan}
\end{figure}

\subsection{Tan's Contact Density}
From our ground-state energy results, we can extract a system property called
the contact density $C$, which has units of $\left[\textrm{length}\right]^4$ and
roughly measures how many pairs of opposite-spin fermions are close
together~\cite{Zwerger2012,Randeria2014}.  In the universal limits of the Fermi
gas, $C$ can be determined via several different approaches via relations
derived by Tan and subsequently expanded upon by others~\cite{
Tan2008a,Tan2008b,Tan2008c,Braaten2008,Werner2009,Zhang2009,Zwerger2012}.  For
example, the contact density determines the tail of the momentum distribution
via $n(k) \sim C/k^4$, and it can also be understood as the conjugate variable
to the interaction strength and thus calculated by the derivative of the
ground-state energy~\cite{Punk2009,Haussmann2009,Zwerger2012,Levinsen2017},
\begin{equation}
    \frac{C}{k_\mathrm{F\uparrow}^4} =
    -\left(5 \pi \right)^{-1}
    \frac{\partial\left( E/\frac{3}{5} N_\uparrow \varepsilon_{\mathrm{F}\uparrow}\right)}
    {\partial \left( 1/\kFup a \right)}
\end{equation}
We calculated the derivative by fitting our data to the functional form 
$E = A-B/C^{1/\kFup a}$.
In Fig.~\ref{fig:tan}, we show our results for the dimensionless contact density
as a function of the interaction parameter $1/\kFup a$ at a representative set
of minority spin concentrations $x$.  We compare CCD to FN-DMC, with the latter
being calculated from the parametrization in Ref.~\onlinecite{Bertaina2009}
based off of the FN-DMC data from Ref.~\onlinecite{Pilati2008}.
We find overall good agreement between the two methods, with CCD predicting a
slightly smaller contact at large interaction strengths. 
For reference, we also include, at each minority spin concentration, the contact
density in the BEC ($1/\kFup a \rightarrow +\infty$) limit to leading order in
$\kFup a$~\cite{ Punk2009,Haussmann2009,Zwerger2012}.

\subsection{Phase Boundary}

\begin{figure}[t!]
  \centering
    \includegraphics[scale=0.9]{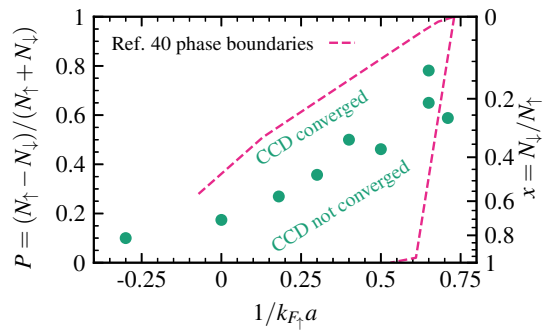}
  \caption{
  Phase diagram of CCD convergence, where the green circles give the maximum
value of $1/\kFup a$ for which CCD converged at each polarization.  The pink
dotted lines give the boundaries for the coexistence region of the partially
polarized normal state with several superfluid states, as determined in
Ref.~\onlinecite{Pilati2008}.
}
  \label{fig:phase}
\end{figure}

As can be seen in Figs.~\ref{fig:energies}(a)-(d), we cannot obtain CCD results
beyond a critical interaction-dependent concentration $x_\mathrm{c}(1/\kFup a)$.
In practice, this occurs when the iterative method used to solve
Eq.~(\ref{eq:t2eqns}) does not converge. This behavior suggests (but does not
guarantee) that Eq.~(\ref{eq:t2eqns}) has no solution.  Similar to the use of
the Thouless criterion for the vertex function~\cite{
Thouless1960,Nozieres1985,Jin2010} or to the presence of multiple solutions in
other nonlinear numerical techniques, we can use this behavior to speculatively
identify a phase boundary for the partially polarized normal state.  In
Fig.~\ref{fig:phase}, we plot the critical polarization,
$P_\mathrm{c} = (1-x_\mathrm{c})/(1+x_\mathrm{c})$, below which the CCD
iteration does not converge for a system of fixed size, $N_k = 1000$.
It is known from previous
works~\cite{Pilati2008,Parish2021,Partridge2006,Shin2006,Radzihovsky2010} that
with increasing interaction strength, the partially polarized normal state
undergoes a first-order phase transition to a region of coexistence with a
superfluid state. At a subsequent second-order phase transition, the coexisting
partially polarized normal state evolves into a coexisting fully polarized
normal state.  These two phase boundaries (determined in previous work) are also
plotted in Fig.~\ref{fig:phase}.  For large polarizations, the CCD convergence
boundary is in good agreement with these phase boundaries.  At smaller
polarizations, we see that the CCD iterations fail to converge in the region of
coexistence of normal and superfluid states.  Fuller insight into the phase
diagram would require a CCD study of the partially and fully polarized
superfluid phases, as performed via FN-DMC in Ref.~\onlinecite{Pilati2008}.

\section{Conclusion}
\label{sec:conc}

To summarize, we have introduced coupled-cluster theory as a promising
computational method for the study of dilute Fermi gases. We have established
the performance of coupled-cluster theory with double excitations (CCD) for the
normal state of polarized three-dimensional gases at a range of interaction
strengths.

Our promising findings motivate a large number of future studies enabled by the
power and generality of the CC framework. In particular, CC theory can be
systematically improved by increasing the number of excitations considered.  It
can be used to calculate one- and two-particle Green's functions directly on the
real frequency axis~\cite{
Koch1990,Nooijen1992,Nooijen1993,Stanton1993,Bhaskaran-Nair2016,Peng2018,Peng2018a,Peng2021},
as done recently for the uniform electron gas with long-range, repulsive Coulomb
interactions~\cite{ McClain2016,Lewis2019}.
This straightforward access to dynamical response functions should be contrasted
with that of most QMC methods, which require analytic continuation. CC methods
can also be applied to nonuniform systems, precluding the local density
approximation for trapped gases, as well as at nonzero temperature~\cite{
Hummel2018,White2018,White2020,Peng2021a,Hermes2015,Harsha2019} and in
non-equilibrium settings~\cite{Dzhioev2015,White2019,Shushkov2019}.
Lastly and perhaps most importantly, CC theory can be formulated with respect to
a BCS reference wavefunction, as opposed to the normal Fermi sea wavefunction
used here, allowing a more accurate study of pairing and superfluidity~\cite{
Lahoz1988,Signoracci2015,Duguet2017err,Henderson2014,Henderson2015,Qiu2019}.
Work along all of these lines is currently in progress.

\section{Data Availability Statement}
\label{sec:das}
The data that support the findings of this study
are openly available at \href{https://github.com/jamescallahan7/jmc_js_tcb_2022}{https://github.com/jamescallahan7/jmc\_js\_tcb\_2022}.

\begin{acknowledgments}
We thank F\'elix Werner for helpful discussions.
J.M.C. was supported by the Department of Defense (DoD) through the National
Defense Science \& Engineering Graduate (NDSEG) Fellowship Program.
J.S. acknowledges support from the National Science Foundation (NSF) Materials
Research Science and Engineering Centers (MRSEC) program through Columbia
University in the Center for Precision Assembly of Superstratic and Superatomic
Solids under Grant No.~DMR-1420634. J.S. also acknowledges the hospitality of
the Center for Computational Quantum Physics (CCQ) at the Flatiron Institute.
We acknowledge computing resources from Columbia University's Shared Research
Computing Facility project, which is supported by NIH Research Facility
Improvement Grant 1G20RR030893-01, and associated funds from the New York State
Empire State Development, Division of Science Technology and Innovation (NYSTAR)
Contract C090171, both awarded April 15, 2010. 
The Flatiron Institute is a division of the Simons Foundation.
\end{acknowledgments}

\end{document}